\documentclass[preprint,12pt]{elsarticle}

\usepackage{graphicx}

\usepackage{amssymb}
\usepackage{amsmath}

\usepackage{lineno}

\newenvironment{mylist}{ 
	\begin{list}{-}{}
}{
	\end{list}
}

\journal{Nuclear Instrument and Methods in Physics Research Section A }

\begin{document}

\begin{frontmatter}



\title{Developing a method for soft gamma-ray Laue lens assembly and calibration}


\author[label1]{Nicolas M. Barri{\`e}re\corref{cor1}}
\author[label1]{John A. Tomsick} 
\author[label1]{Steven E. Boggs}
\author[label1]{Alexander Lowell}
\author[label2]{Colin Wade}
\author[label1]{Max Baugh}
\author[label3]{Peter von Ballmoos}
\author[label4]{Nikolay V. Abrosimov}
\author[label2]{Lorraine Hanlon}

\address[label1]{Space Sciences Laboratory, 7 Gauss Way, University of California, Berkeley, CA 94720-7450, USA} 
\address[label2]{School of Physics, University College Dublin, Dublin 4, Ireland}
\address[label3]{Institut de Recherche en Astrophysique et Planetology, UMR 5277,  9 av. du Colonel Roche, 31028 Toulouse,  France}
\address[label4]{Institute for Crystal Growth, Max-Born-Str.2 D-12489 Berlin, Germany}

\cortext[cor1]{Corresponding author: N.B. \\E-mail: barriere {\it at}  ssl.berkeley.edu}

\begin{abstract}
Laue lenses constitute a promising option for concentrating soft gamma rays with a large collection area and reasonable focal lengths. In astronomy they could lead to increased telescope sensitivity by one to two orders of magnitude, in particular for faint nuclear gamma-ray lines, but also for continua like hard X-ray tails from a variety of compact objects. Other fields like Homeland security and nuclear medicine share the same need for more sensitive gamma-ray detection systems and could find applications for gamma-ray focusing optics.
There are two primary challenges for developing Laue lenses: the search for high-reflectivity and reproducible crystals, and the development of a method to accurately orient and fix the thousands of crystals constituting a lens. In this paper we focus on the second topic. We used our dedicated X-ray beamline and Laue lens assembly station to build a breadboard lens made of 15 crystals. This allowed us to test our tools and methods, as well as our simulation code and calibration procedure. Although some critical points were identified, the results are very encouraging, with a crystal orientation distribution lower than $10''$, as required to build a Laue lens telescope dedicated to the study of Type Ia supernovae (30-m focal length). This breadboard lens represents an important step towards raising the technology readiness level of Laue lenses.
\end{abstract}


\end{frontmatter}


\section{Introduction}
\label{sec:intro} 

Observations of the sub-MeV gamma-ray sky enable direct glimpses of fundamental physics processes involving conditions that are not reproducible in the laboratory, such as extreme magnetic fields up to $10^{15}$ G near magnetized neutron stars or extreme gravitational fields near black holes. However, observations are hampered by the limited sensitivity of current telescopes. High instrumental background in detectors is the main problem, and building bigger detectors is not a viable solution as the sensitivity only (roughly) scales with the square root of the detector surface area. A Laue lens telescope (LLT) allows the decoupling of the collecting area from the detector area, dramatically increasing the signal-to-noise ratio and hence the sensitivity. The benefits of focusing high-energy radiation was recently demonstrated once again with NASA's observatory {\em NuSTAR} extending the focused bandpass to 80 keV \cite{harrison.2013ly} (the maximum was previously $\sim$12 keV). {\em NuSTAR} is providing an entirely new view of the hard X-ray sky with unprecedented sensitivity.

One topic that would benefit from the advent of a LLT is the study of the Type Ia supernovae (SNe Ia). SNe Ia are used as a cosmological standard candle to determine extra-galactic distances, which has led to the astonishing result that the expansion of the Universe is accelerating, implying the existence of dark energy \cite{riess.1998uq, perlmutter.1999kx}. However, we do not understand why SNe Ia luminosities can be normalized \cite{1993ApJ...413L.105P}, which is related to our lack of understanding of the progenitor system and the physics of the explosion. The spectroscopy and light curve of the line at 847 keV emitted by the decay chain of $^{56}$Ni, which is massively synthesized in SNe Ia, would discriminate between the currently competing models \cite{branch.1995uq}. A LLT, as featured in the {\em DUAL} mission proposal \cite{von-ballmoos.2011kx}, could reach a sensitivity of $2 \times 10^{-6}$ ph/s/cm$^2$ (3 $\sigma$, 1 Ms)  for a 3\% broadened line at 847 keV, enabling detections of a dozen events each year out to $\sim$40 Mpc and providing a breakthrough in our understanding of their physics \cite{barriere.2010fk}.

Another topic is the study of the electron-positron annihilation radiation at 511 keV. This line has been observed for more than 30 years from the Galactic center \cite{leventhal.1978ay}, yet it is still unclear whether known sources can account for all of the 10$^{43}$ positrons that annihilate every second in the Galactic bulge \cite{knodlseder.2005rc}. New observational clues are needed, requiring both improved sensitivity and angular resolution. A LLT could probe small sky regions to check for structure in the emission and probe some candidate source types, like X-ray binaries. 

Other objectives include the study of the emission mechanisms in blazars and active galactic nuclei \cite[e.g.][]{abdo.2010fk} and the physics of stellar mass black holes in binary systems \cite[ e.g.][]{grove.1998wt} through the observation of their emission in energy bands within the 100 keV - 1 MeV domain.

Laue lenses are an emerging technology based on crystal diffraction that enables soft gamma-ray focusing. The advent of this optic would highly benefit hard X-ray and soft gamma-ray astrophysics, along with other fields. For instance, homeland security and nuclear medicine share the same need for more sensitive gamma-ray detection systems. A Laue lens offers a narrow field of view (typically of $\sim$10$'$), and can be designed to focus in a narrow energy bandpass, which can turn into advantages for applications where background is an issue and spatial resolution is required (for instance looking for fluorescence lines from a target material, activated by a gamma-ray beam \cite{albert.2011fk}).

UC Berkeley's Space Science Laboratory (SSL) joined the effort to develop Laue lenses in 2010, building upon the experience accumulated over the past 20 years at the IRAP (Toulouse, France) \cite[][]{ballmoos.1994oq, kohnle.1998wj, halloin.2003xy}. A dedicated X-ray beamline was completed in Spring 2011, which then allowed the development of an assembly method. The challenge of making a scientifically exploitable Laue lens can be divided in two topics: finding efficient crystals for diffraction, and assembling them accurately enough into a lens. The study and development of crystals for a Laue lens application has been on-going for nearly a decade, resulting in the identification of the best crystals for each energy within the ~100 keV - 1 MeV band \cite{Barriere:he5432, rousselle.2009kx, barriere.2011fk, ferrari.2012fk}. The crystal selection is not discussed in this paper. Instead, here we focus on the second aspect, their assembly into a Laue lens. We report on the assembly tools and method that were used to build a breadboard lens made of 15 crystals, and on the calibration procedure and results. This test confirms our ability to reach the crystal orientation accuracy and the packing factor required to build an efficient Laue lens with a focal length of several tens of meters, as required for Type Ia supernovae study for instance \cite{barriere.2010fk}.

This paper is organized as follows: The concept of Laue lenses is reviewed in section \ref{sec:Lauelens}. Section \ref{sec:reference} introduces the coordinate system we used during this work. Section \ref{sec:orientation} presents the crystal orientation requirements for a Type Ia SNe LLT similar to what was presented in Ref. \cite{von-ballmoos.2011kx}, which sets the objectives for the prototype we assembled. Then we enter the heart of the paper with the description of the prototype in section \ref{sec:LaueLens} and the assembly method in section \ref{sec:assemblymethod}. The characterization of the lens prototype is presented in section \ref{sec:characterization}, and finally the conclusions are presented in section \ref{sec:conclusions}.

\section{Principle of a Laue lens} 
\label{sec:Lauelens} 

\begin{figure}[htbp]
\begin{center}
\includegraphics[width=0.66\textwidth]{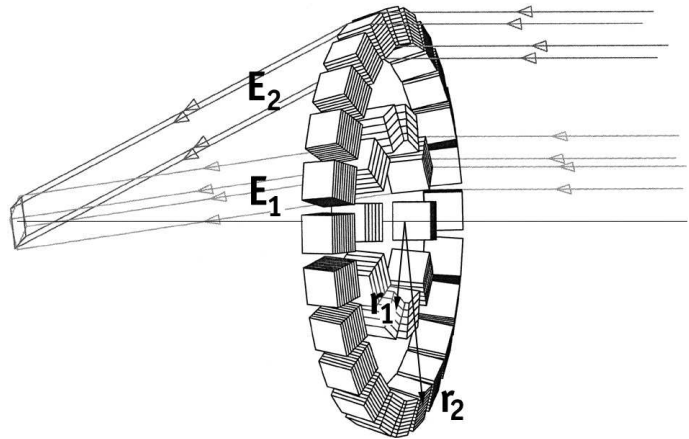}
\caption{Sketch of a Laue lens made of crystals arranged in concentric rings.  If the same crystal material and reflection (which determines the d-spacing of the crystalline planes) are used for the two rings of radii $r_1$ and $r_2$, then $E_1 > E_2$, which allows covering a large bandpass. However, if the product of the  d-spacing by the radius is constant from ring to ring, the energy diffracted is constant, which allows building up effective area in a narrow energy range.\cite[e.g.][]{halloin.2005fk}}.
\label{fig:lensconcept}
\end{center}
\end{figure}

A Laue lens is a concentrator working in the domain of the hard X-rays and soft gamma rays, from $\sim$100 keV to  $\sim$1 MeV. It is based on Bragg diffraction in Laue geometry ({\it i.e.} the rays go through the crystal) of a large number of crystals arranged such that they all diffract towards a common point on the focal plane (Figure \ref{fig:lensconcept}) \cite[see e.g.][]{lund.1992ft, halloin.2005fk, frontera.2010fk}. The crystals can be laid out either in an Archimedean spiral or in concentric rings. The typical cross sectional size of crystals considered for Laue lenses ranges between $4 \times 4$ mm$^2$ \cite{roa.2001fk} to $15 \times 15$ mm$^2$ \cite{barriere.2005fp}. 

In the classical design, each crystal deviates a fraction of the beam without concentrating it. Thus, smaller crystals produce a smaller point spread function (PSF), although at the cost of a larger complexity (larger number to obtain a given collecting area, more difficult to manipulate and orient). Alternatively, a group in Ferrara (Italy) is developing an interesting concept of curved crystals where the diffracted spot of the crystal (hereafter referred to as the crystal's footprint) is smaller than the crystal itself \cite{camattari.2013fk}.

Perfect crystals are not suitable for Laue lens applications as they behave as monochromators. Even for the case of a Laue lens dedicated to the observation of a given nuclear line on the ground\footnote{As opposed to an astrophysical context.}, crystals with a spread in the orientation of their planes are more efficient than perfect crystals. This is due to the fact that the source is at finite distance, implying that the beam hitting each crystal is diverging. Thus, a crystal can diffract over its full volume only if it presents to the source a bandpass at least matching the angle subtended by its cross-sectional area. 
Mosaic crystals are the most common mono-crystalline non-perfect crystals. Their bandpass is created by small defects in their crystal lattice\footnote{The term {\it mosaic crystal} comes from the fact that they are well modeled by a juxtaposition of tiny, perfect crystals, slightly disoriented with respect to each other, as proposed by Darwin \citep{darwin.1914dz, darwin.1922zg}.}. Alternatively, crystals with curved diffracting planes (CDP crystals) can yield higher reflectivity, however they are more difficult to produce \cite{Barriere:he5432}. 

Most Laue lens projects require several thousands of crystals, with focal lengths of several tens of meters. The consequences are twofold: On the one hand, the time devoted to fix each crystal should be as short as possible; and on the other hand, the crystals should be oriented with high accuracy in order to keep the PSF as small as possible. For a given lens design\footnote{We use the term {\it lens design} to refer to a given combination of crystals (material, orientation, bandpass), and tile size, and their configuration in the lens (radii, and filling factor of each ring).}, the effective area scales with the number of crystals collecting the signal, which is why the crystals should be packed as densely as possible.

\section{Lens reference system} 
\label{sec:reference}

\begin{figure}[htbp]
\begin{center}
\includegraphics[width=0.66\textwidth]{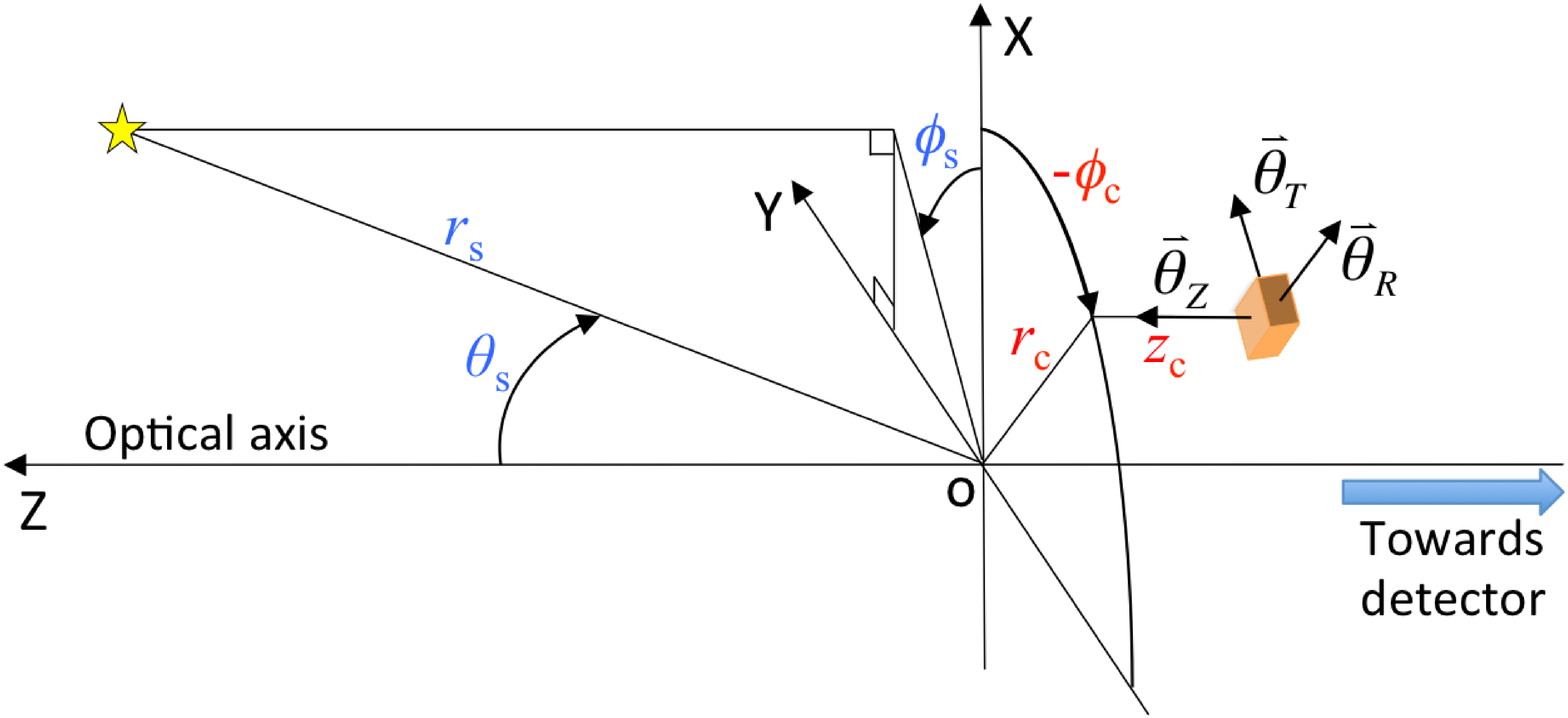}
\includegraphics[width=0.32\textwidth]{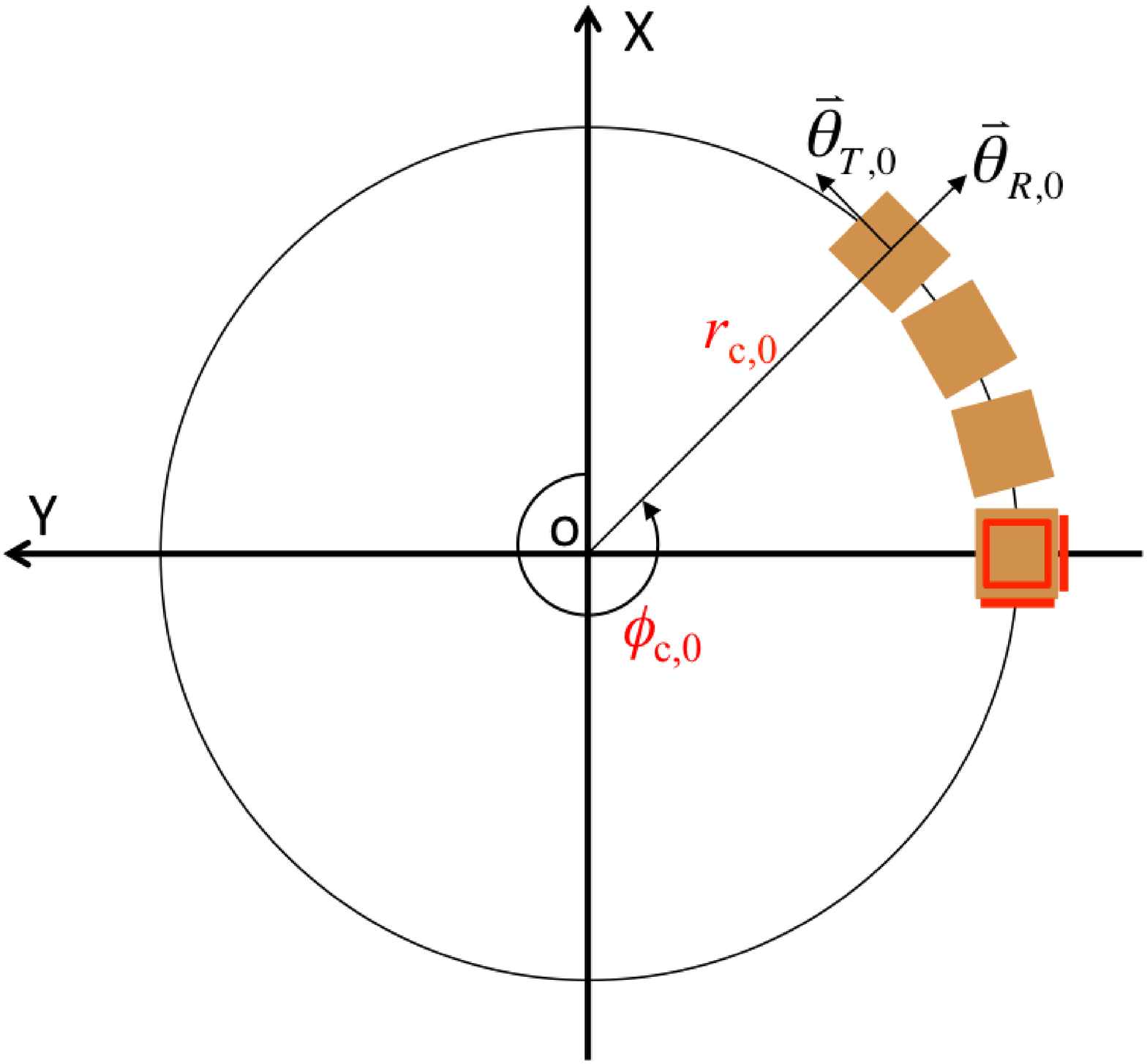}
\caption{{\it Left:} Lens reference system. See text for details. {\it Right:} Sketch of four crystals on the lens frame, the fourth being glued. }
\label{fig:reference}
\end{center}
\end{figure}

Figure~\ref{fig:reference} shows the reference system we use for this study. The lens plane is XoY, and oZ defines the optical axis, with the rays propagating towards -Z. 

The source position $[r_s, \theta_s, \phi_s]$ is defined in spherical coordinates based on the lens reference system. The crystal positions $[r_c, \phi_c, z_c]$ are defined in cylindrical coordinates based on the lens reference system. Each crystal orientation is determined with respect to axes defined by the crystal position on the lens. The orientation is defined by rotations about the radial, tangential, and optical axes, noted $\vec{\theta_R}$, $\vec{\theta_T}$, and $\vec{\theta_Z}$. In the ideal case and for an on-axis source at infinity, $\theta_T$ is the Bragg angle and the two other angles are null.

\section{Crystal orientation requirements}
\label{sec:orientation}

\begin{figure}[htbp]
\begin{center}
\includegraphics[width=0.48\textwidth]{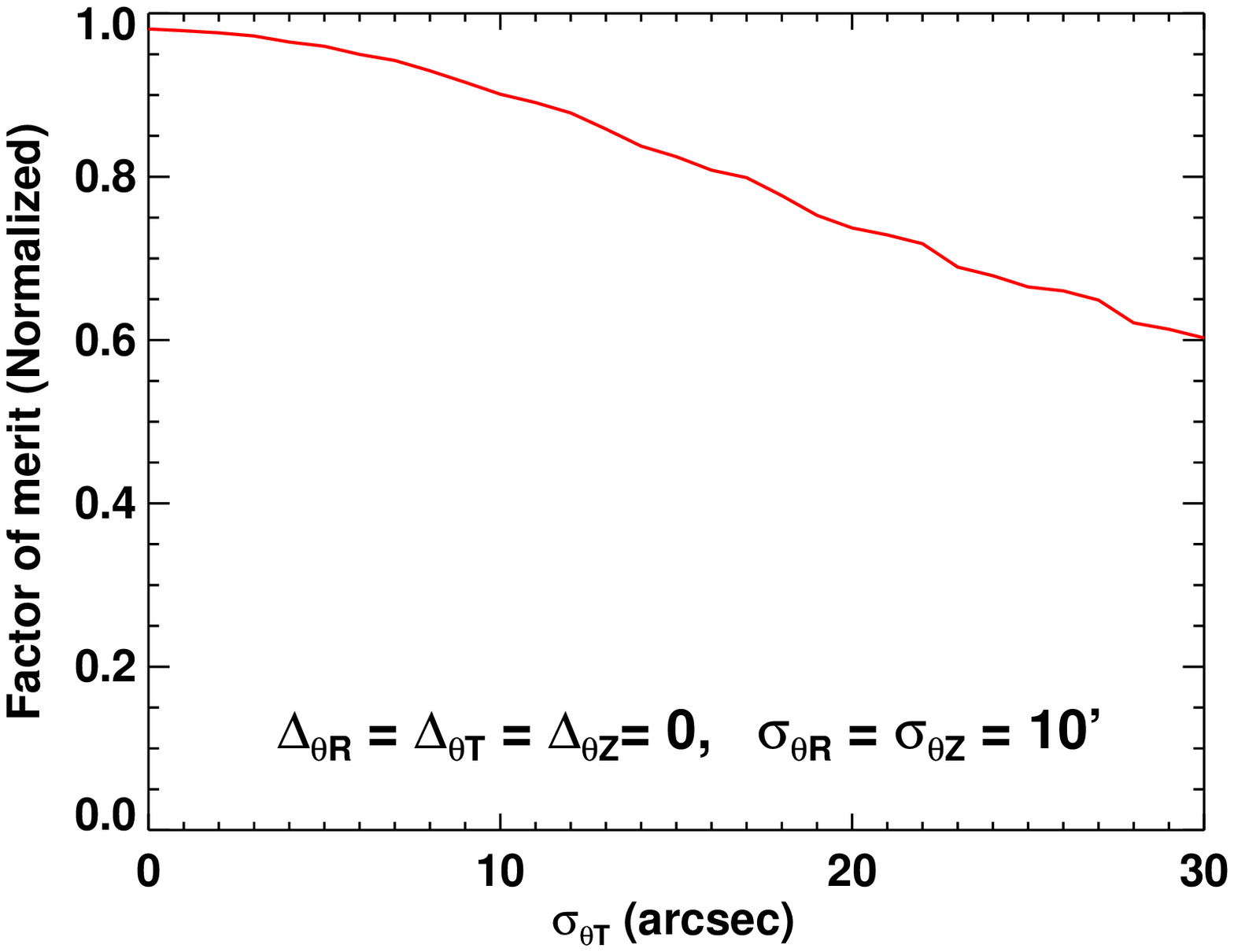}
\includegraphics[width=0.48\textwidth]{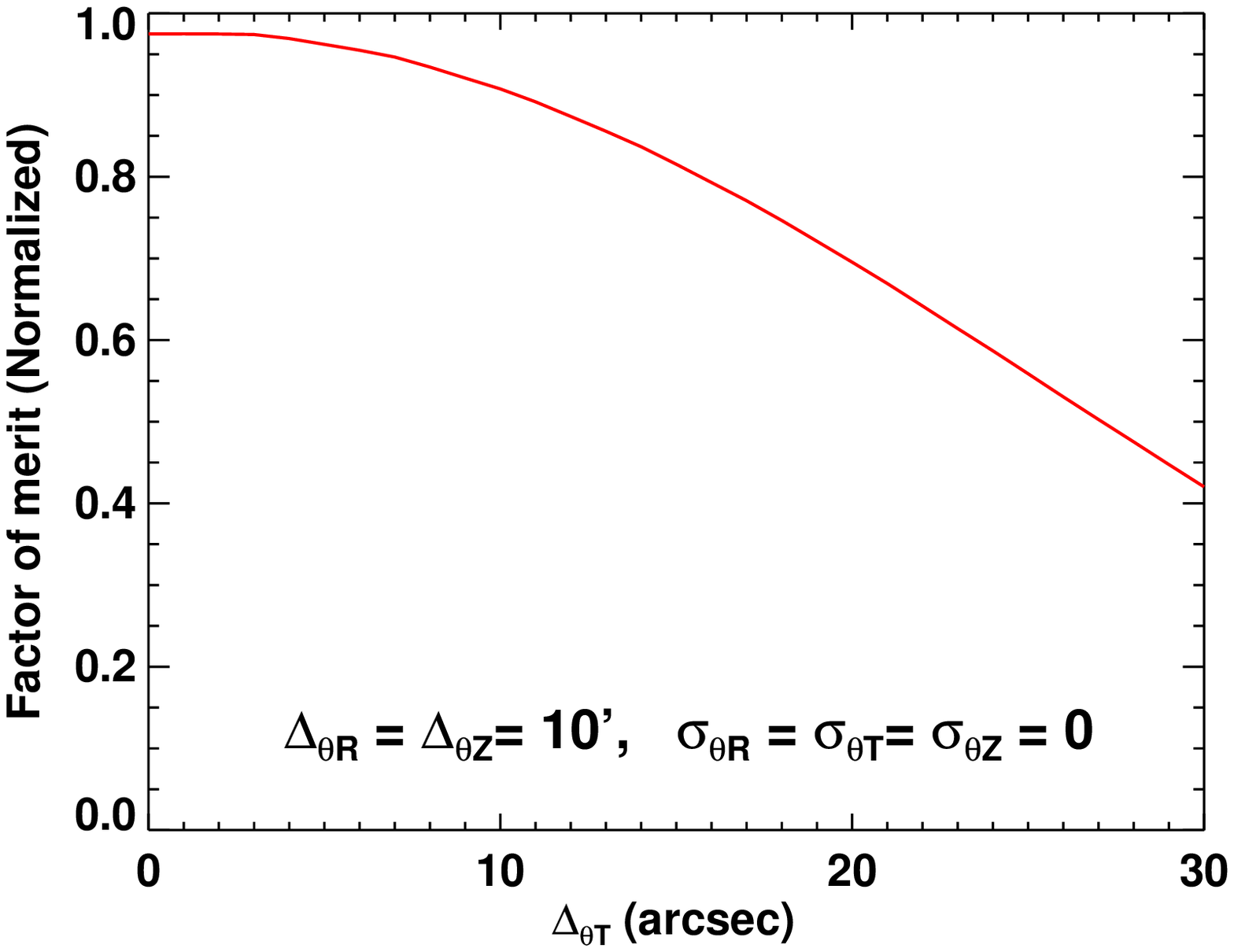}
\caption{Figure of merit (normalized) versus Bragg angle misalignment standard deviation ($\sigma_{\theta T}$, left panel) and Bragg angle misalignment offset ($\Delta_{\theta T}$, right panel). In the left panel, the offsets are fixed to 0, and the standard deviation $\sigma_{\theta R}$ and $\sigma_{\theta Z}$ are fixed to $10'$. In the right panel, the standard deviations are fixed to 0 and the offsets $\Delta_{\theta R}$ and $\Delta_{\theta Z}$ are fixed to 10'. In both cases, the simulations are done for a 30-m focal length lens made of $10\times10$ mm$^2$ mosaic Ag 111 crystals focusing at 850 keV.}
\label{fig:orientation}
\end{center}
\end{figure}

In order to specify the crystal orientation accuracy requirements, one needs a measure of the impact of crystal angular offset. The only relevant figure is the sensitivity of the telescope. Assuming that the instrumental background is uniform in the focal plane, the sensitivity of the telescope is proportional to the following figure of merit (FoM):
\begin{eqnarray}
FoM = {{A_{eff} \,\epsilon_{PSF}} \over \sqrt{A_{PSF}} }
\end{eqnarray}
where $A_{eff}$ is the lens effective area, and $A_{PSF}$ is the area covered by the fraction $\epsilon_{PSF}$ of the PSF (the choice of $\epsilon_{PSF}$ is discussed below). This FoM is expressed in cm, but it is more relevant to normalize it by the value obtained for a lens made of ideally oriented crystals. 

The distribution of angular misalignment is considered Gaussian and is described by two parameters: its standard deviation, noted $\sigma_{\theta R}$, $\sigma_{\theta T}$, $\sigma_{\theta Z}$, and the offset between the center of the distribution and the nominal angle, noted $\Delta_{\theta R}$, $\Delta_{\theta T}$, $\Delta_{\theta Z}$, respectively for the 3 angles $\theta_R$, $\theta_T$, and $\theta_Z$. The standard deviation of the distribution affects the width of the energy bandpass and the size of the PSF. The offset of the distribution affects the energy diffracted and the focal length. A crystal ring with a non-zero offset does not focus at the proper focal length, implying a size increase of the PSF.

We decouple the standard deviation from the offset, and we first investigate the effects of the former. We calculated the FoM for a simulated lens made of $10\times10$ mm$^2$ Ag 111 crystals arranged in a single ring focusing at 850 keV with a 30-m focal length. Each crystal has a uniform mosaicity of 45$''$ and a mean crystallite size of 100 $\mu m$.  In these simulations, all the offsets are kept to 0, which means that the mean orientations are nominal along the three axes. Figure~\ref{fig:orientation} shows the FoM as a function of $\sigma_{\theta T}$, the standard deviation of misalignment of the Bragg angle, with both $\sigma_{\theta R}$ and $\sigma_{\theta Z}$ set to $10'$. For each value of $\sigma_{\theta T}$, the lens' PSF and effective area are simulated and the FoM is derived using the combination of $A_{PSF}$ and $\epsilon_{PSF}$ that maximizes it. 

The FoM is most sensitive to the misalignment of the Bragg angle. One can see in Figure~\ref{fig:orientation} (left panel) that for $\sigma_{\theta R}=\sigma_{\theta Z}=10'$ and $\sigma_{\theta T}=0$ the sensitivity loss is merely 2\%; however the FoM drops by 10\% for $\sigma_{\theta T} = 10''$. 

One also needs to account for the offset of the distribution along each axis. Another set of simulations was performed. At first, the standard deviations were null, only the offset were varied. This showed that only $\Delta_{\theta T}$ really matters; for $\Delta_{\theta R}= 10'$, $\Delta_{\theta Z}=10'$ and $\Delta_{\theta T}=0$, the FoM drops by only 2\%, however the FoM drops by 9\% if $\Delta_{\theta T}=10''$ (Figure \ref{fig:orientation}, right panel).

Performing more simulations combining both offset and standard deviation of misalignment for the 3 axes, we derive the orientation requirements in order to limit the FoM loss to 10\%. We obtain the following requirements:
\[
\sigma_{\theta R} \leq 7' \;,\;
\sigma_{\theta T} \leq 10'' \;,\;
\sigma_{\theta Z} \leq 7' \;,\;
\Delta_{\theta R} \leq 5' \;,\;
\Delta_{\theta T} \leq 4'' \;,\;
\Delta_{\theta Z} \leq 5'
\]

\section{Description of the lens}
\label{sec:LaueLens} 
\begin{figure}[htbp]
\begin{center}
\includegraphics[width=0.95\textwidth]{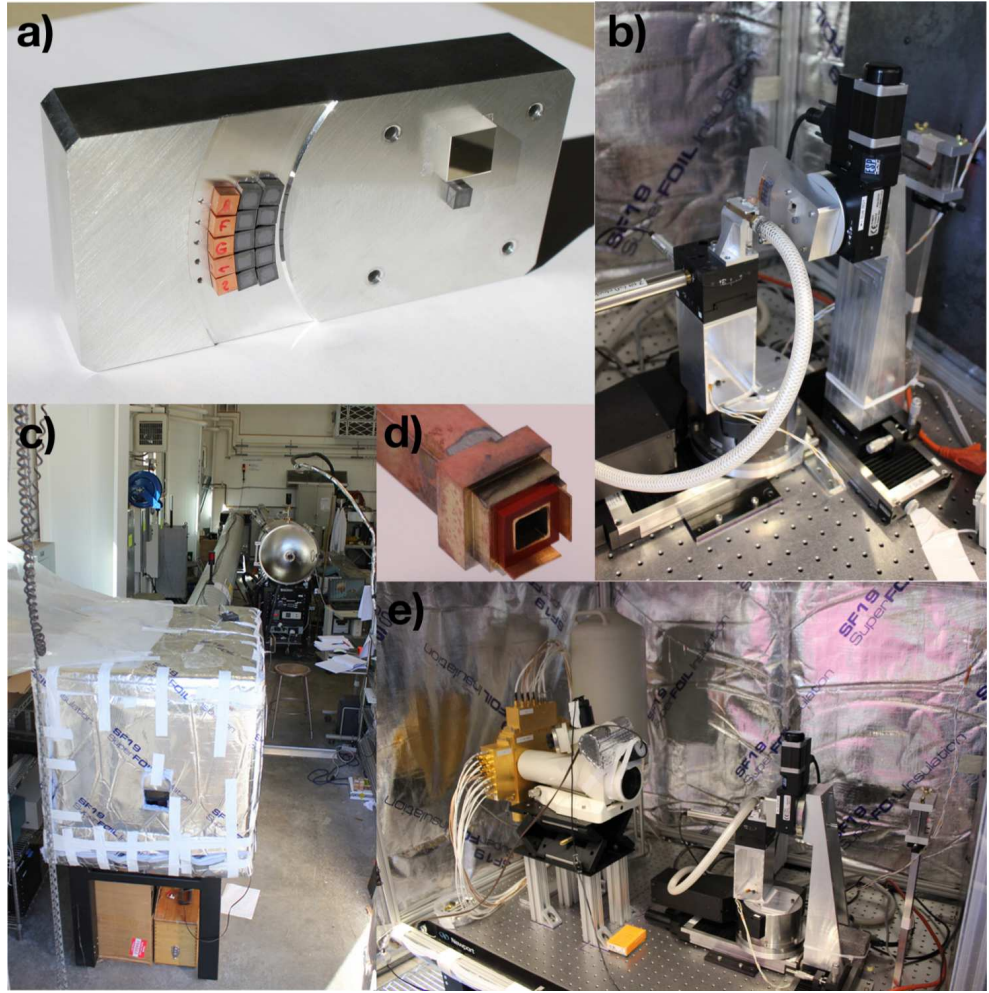}
\caption{{\it a)} Laue lens prototype. {\it b)} Substrate and crystal towers. The slits defining the beam to 4 $\times$ 4 mm$^2$ are also visible. {\it c)} The beamline at SSL is setup in the high bay. In the foreground is the thermally insulated LLAS. {\it d)} Close up on the tip of the crystal holder. The two ledges are visible on the right hand side and at the bottom. The red square is a rubber O-ring.  {\it e)} Full view of the LLAS. The station is setup on a $30 \times 48$ inches$^2$ Newport table.}
\label{fig:photos}
\end{center}
\end{figure}

The prototype lens is composed of 5 Cu crystals and 10 Si crystals arranged in 3 sections of concentric rings (Figure~\ref{fig:photos}a), as detailed in Table \ref{tab:lensdetails}. The Cu crystals were produced at the Institut Laue Langevin (Grenoble, France), and the Si crystals were produced at the Institute for Crystal Growth (IKZ, Berlin, Germany). The crystal dimensions are $ 5 \times 5 \times 3 $ mm$^3$ and the crystal interspacing is 0.2 mm at the closest point (distance between the innermost corners of two neighboring crystals). The lens is designed to focus the beam of our X-ray generator (XRG) placed at $r_S=12.49$ m with a focal length of $f=1.5$ m. 

The crystals are glued on the substrate, thus the orientation relies on the glue bond line. The lens substrate is made of aluminum, with slopes (portions of cones) following the $\theta_T$ angle of the crystals in order to keep the glue bond line nearly parallel\footnote{Crystals are usually cut within $10'$ of the required orientation, which results in some uncertainty in the bond line shape. This is why we can not rely on the faces of the crystal tiles for the orientation.}. The substrate's back side is milled out to reduce passive material, its thickness is about 2~mm. It features holes at the center of each crystal site in order to inject the glue from the back side.

\begin{table}[htdp]
\begin{small}
\begin{center}
\begin{tabular}{|c|c|c|c|c|c|}
\hline
Ring 	& Reflection	& Radius 	& $\theta_T$ 	& Bragg angle  & Energy \\
\#		&  $(hkl)$		& (mm) 	& ($^{\circ}$)	& ($^{\circ}$)	& (keV)	\\
\hline
\hline
0 		& Si 111		& 52.0	& 0.8734		& 1.1120	 	& 101.878 \\
\hline
1		& Si 111		& 57.2	& 0.9607		& 1.2231		& 92.625  \\
\hline
2		& Cu 111		& 62.4	& 1.0479		& 1.3342		& 127.565 \\
\hline
\end{tabular}
\end{center}
\end{small}
\caption{Nominal orientation for each crystal ring of the prototype lens. The difference between the Bragg angle (the incidence angle) and $\theta_T$ is due to the fact that the source is at finite distance. \label{tab:lensdetails}}
\end{table}%

\section{Assembly method} 
\label{sec:assemblymethod}

\begin{figure}[htbp]
\begin{center}
\begin{minipage}[c]{.46\linewidth}
      \includegraphics[width=0.8\textwidth]{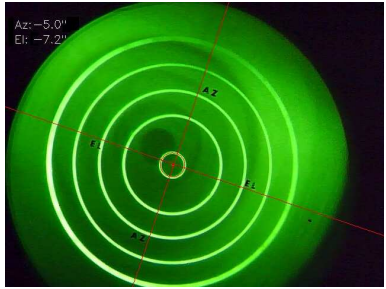}
\end{minipage} \hspace{-0.5cm}
\begin{minipage}[c]{.46\linewidth}
      \includegraphics[width=\textwidth]{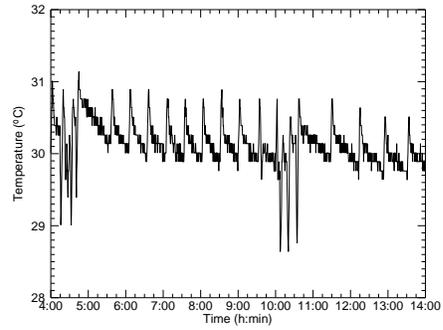}
\end{minipage}
\caption{{\it Left:} autocollimator shot taken with the webcam and processed. The crosshair has been identified as well as the center of the bright green concentric circles, which allows the determination of the angular distance between the center of the circles and the two arms of the cross. {\it Right:} Evolution of the temperature in the Laue lens assembly station over ten hours.}
\label{fig:temperaure}
\end{center}
\end{figure}

\subsection{The Laue lens assembly station}
\label{sec:LLAS}
The Laue lens assembly station (LLAS) that we developed at SSL is placed at the end of a 12-m long X-ray beamline using a micro-focus (0.8 mm) XRG operated at 150 kV and 450 $\mu$A (Figure~\ref{fig:photos}c). This beamline was already presented in Ref. \cite{barriere.2011fk} and has not changed since then, however, a number of changes were implemented in the LLAS (Figure~\ref{fig:photos}e). The LLAS is composed of the following elements, in the order of the beam propagation:

\begin{mylist}
\vspace{-0.3cm}
\item A set of slits defining the beam to $4 \times 4$ mm$^2$ (visible in the right hand side of Figure~\ref{fig:photos}b). The beam position was set prior to the lens assembly and was not touched after.
\vspace{-0.3cm}
\item The lens aluminum substrate held by a stack of stages (Figure~\ref{fig:photos}b): a translation perpendicular to the beam (along oY) to change the radius, and a rotation stage (axis oZ) to change the azimuthal position. The rotation axis of this stage defines the axis of symmetry of a ring and thus its optical axis. In addition to these two stages, a manual tilt and rotation stage was inserted on top of the translation stage to orient the substrate with respect to the beam.
\vspace{-0.3cm}
\item The crystal to be glued, held by the crystal holder (which uses vacuum suction to maintain the crystal) at the top of a stack of stages allowing 3-axis rotations. Given our setup (Figure~\ref{fig:reference}), the axes $\vec{\theta_R}$, $\vec{\theta_T}$, $\vec{\theta_Z}$ of the crystal match with the directions -oY, oX, oZ, respectively.  The stack is mounted on a translation stage (along oZ) to bring the crystal against the substrate once its orientation is correct (Figures \ref{fig:photos}b and \ref{fig:photos}d).
\vspace{-0.3cm}
\item The detector (visible in Figure~\ref{fig:photos}e). We have been using a planar cross-strip high-purity germanium detector measuring $38 \times 38$ mm$^2$ divided in $19 \times 19$ voxels of $2 \times 2$ mm$^2$ \cite{amrose.2003uq}. This camera is a prototype for the Nuclear Compton Telescope \cite{bandstra.2011uq}. It allows the extraction of the spectrum from any number of voxels, with a spectral resolution of 1.3 keV at 122 keV.
\vspace{-0.3cm}
\end{mylist}

In our setup, the crystal being glued is mounted at the tip of the crystal holder, which is centered on the beam. The crystals are glued at $\phi = 270^{\circ}$, diffraction occurring in the horizontal plane (YoZ, see Figure{fig:reference}). The crystal position and orientation is defined by the plane of the crystal holder, represented as a red square, and by the two ledges of the crystal holder, shown as two red segments in the right panel of Figure \ref{fig:reference} (see also Figure~\ref{fig:photos}d). The lens is moved to set the radius $r_c$ and azimuthal angle $\phi_c$ of the new crystal. The radius $r_c$ is controlled by an oY-translation stage holding the lens, and the azimuthal position of the crystal is controlled by an oZ-rotation stage that holds the lens substrate. One sees in the right panel of Figure~\ref{fig:reference} that the angles $\theta_R$ and $\theta_T$ are controlled by the plane of the crystal holder, while $\theta_Z$ is controlled by the two ledges of the crystal holder.

In 2011, our first attempt to glue crystals with angular precision lower than $10''$ taught us that controlling the temperature is a necessity \cite{barriere.2011fk}. The LLAS was thermally insulated and a commercial thermostat coupled to a small fan heater was used to maintain the temperature around 30$^{\circ}$C (above the maximum temperature observed in the room). Despite this low-cost system, the temperature was maintained within a range of $1.3^{\circ}$C during the gluing (Figure~\ref{fig:temperaure}). The big drops in temperature happen when the doors of the LLAS are opened either to setup a new crystal on the holder or to inject the glue. One can see that the total duration for crystal setup and orientation and glue injection was of the order of 30 minutes, followed by about 5 hours of curing time.

We use a Davidson Optronics D-656 autocollimator (visible in Figure~\ref{fig:photos}e) with arc-second precision to monitor the orientation of the aluminum substrate (see section \ref{sec:OrientSub}). We automated the reading of the autocollimator by adding a webcam on the eyepiece and developed software that analyzes the image to return the azimuth and elevation of the bullseye center with respect to the optical axis of the instrument (Figure \ref{fig:temperaure}). This was key to the realization of this project as it allows  the orientation of the lens substrate to be monitored remotely, without disturbing the temperature.

\subsection{Orienting the substrate}
\label{sec:OrientSub}

In our setup, the optical axis of a crystal ring is given by the rotation axis of the substrate, which needs to be set for each ring (see below). If this alignment is not done properly, different crystal rings may have different optical axes, leading to an overall PSF increase. 

The substrate is correctly oriented when its rotation axis points towards the source. The orientation is set by using the so-called rotating crystal method, involving a crystal\footnote{We used a perfect Si 111 crystal of $5 \times 5$ mm$^2$ as central crystal.} glued at the rotational center of the substrate. The peak energy of the beam diffracted by this crystal is measured for different azimuthal angles. When the rotation axis points at the source, the peak of the energy diffracted by the crystal is constant for any azimuthal angle. The accuracy of this method was limited in our case to about $\pm 5''$ by the accuracy of the tilt stage (manual tilt and rotation stage Newport 36). Further details are presented in section \ref{sec:pointingerror}.

Once the initial orientation is done, the substrate is moved to bring the first crystal site in the beam. The autocollimator then becomes the only way to track the orientation of the substrate, as the central crystal is no longer in the beam. Bringing a crystal site in the beam consists in two steps: the substrate is translated along oY to bring the desired radius in the beam ({\it i.e.} to set $r_c$), and it is then rotated about oX to bring its rotational axis to point again at the source. The autocollimator is used to measure and correct for the wobble induced by the oY translation stage, and control the rotation about oX as it is done with a manual stage. The autocollimator was also used to monitor the orientation of the substrate while the crystals of a given ring were glued. Despite the poor thermal control, we found that the substrate orientation was very stable with time, so we had to re-orient it only when we were changing $r_c$.

\subsection{Orienting and gluing crystals}
\label{sec:OrientGlueCryst}
The process to glue a crystal is the following. The crystal is setup at the tip of the holder, the two little ledges defining the angles $\theta_Z$ and the plane of the holder defining $\theta_R$ and $\theta_T$ (Figure~\ref{fig:photos}{\it d}). The crystal holder had previously been oriented by using a corner cube and the autocollimator, so its suction plane was perpendicular to the beam and the vertical ledge was vertical (the beam being horizontal). We estimate the error on these angles to be less than $5'$. We relied on the crystal external faces for $\theta_R$ and $\theta_Z$, which means an orientation accuracy of $\sim$$10'$ (based on the cutting specifications). 

The crystal is first kept 5~mm in front of the face of the substrate for coarse orientation, it is then brought to $\sim$80~$\mu$m of the substrate for fine orientation and gluing. The Bragg angle is set using our $0.3''$ repeatable oX rotation stage ($\vec{\theta_T}$) to obtain the desired energy diffracted on the camera. When the Gaussian fit of the diffracted peak indicates a misalignment lower than $3''$, the glue is injected through the hole in the substrate, using a syringe. The glue, a two-part epoxy (MasterBond EP30-2) meets NASA low outgassing specifications and has a very low shrinkage upon cure, $3 \times 10^{-4}$ mm/mm. It reaches 85\% of its strength after 12h and its ultimate strength is attained after 5--7 days. To speed up the process, the crystal holder was retracted after $\sim$5 h of glue curing time. 

We are currently developing a method to avoid relying on the faces of the crystals for setting $\theta_Z$ and $\theta_R$, which would relax the cutting accuracy requirement (thus lowering the cutting cost). The method uses the crystalline planes perpendicular to $\vec{\theta_T}$ and rocks the crystal about $\vec{\theta_R}$ to diffract in the vertical plane. Finding the diffraction peaks above (1) and under (-1) the horizontal gives the orientation of the crystalline planes and allows adjusting $\vec{\theta_R}$ parallel to the beam. Then the crystal is rotated by $\pi/2$ about  $\vec{\theta_T}$ and the same procedure is repeated for $\vec{\theta_Z}$. This method is possible with our setup as the crystal is placed on the rotation axis of $\theta_T$ (the bottom rotation stage in picture \ref{fig:photos}b), which allows a $\pi/2$ rotation while keeping the crystal in the beam.

\section{Characterization of the lens}
\label{sec:characterization}

The characterization was done using full flood illumination (we use a 5-mm thick lead mask with an aperture of $2.54 \times 2.54$ cm$^2$). The lens substrate is oriented to point at the source and the 15 crystals are centered in the aperture. The beam is strongly diverging in this configuration, but this is fine because the lens was designed for a source at finite distance, at $r_S=12.49$ m, which is the case here. 

\begin{figure}[t]
\begin{center}
\includegraphics[width=0.47\textwidth]{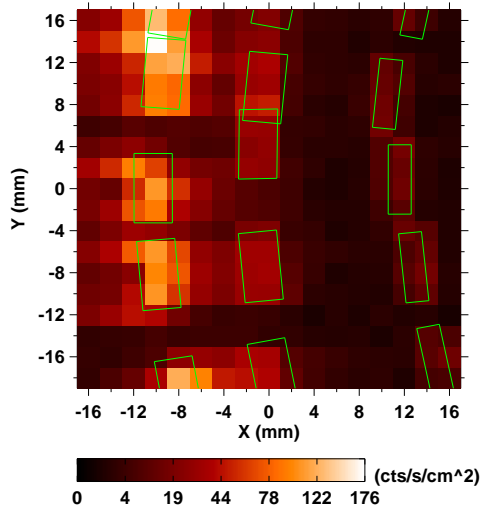}
\caption{Out-of-focus image acquired with our germanium cross-strip camera. The source is on axis and the detector is placed 4 m behind the lens. The green rectangles are the simulated footprint of the 15 crystals projected onto the detector plane, accounting for orientation errors. One can see that our model of the lens and ray-trace code allow for a good reproduction of the observed pattern.}
\label{fig:PSFoutoffocus4m}
\end{center}
\end{figure}

The main measurement was performed with the focal plane out of focus, 4~m behind the lens. This allows blowing up the focal point to reveal each individual crystal footprint (Figure~\ref{fig:PSFoutoffocus4m}). This measurement serves two purposes: Firstly, the energy diffracted by each crystal can be measured individually, yielding an accurate measurement of the Bragg angle ($\theta_T$) misalignment. Secondly, the position of the footprint can be used to infer the $\theta_Z$ misalignment.

The characterization of the lens was done one month after its assembly, to let enough time for the epoxy to stabilize. 


\subsection{Errors on the Bragg angle}

\begin{figure}[p]
\begin{center}
\includegraphics[width=0.8\textwidth]{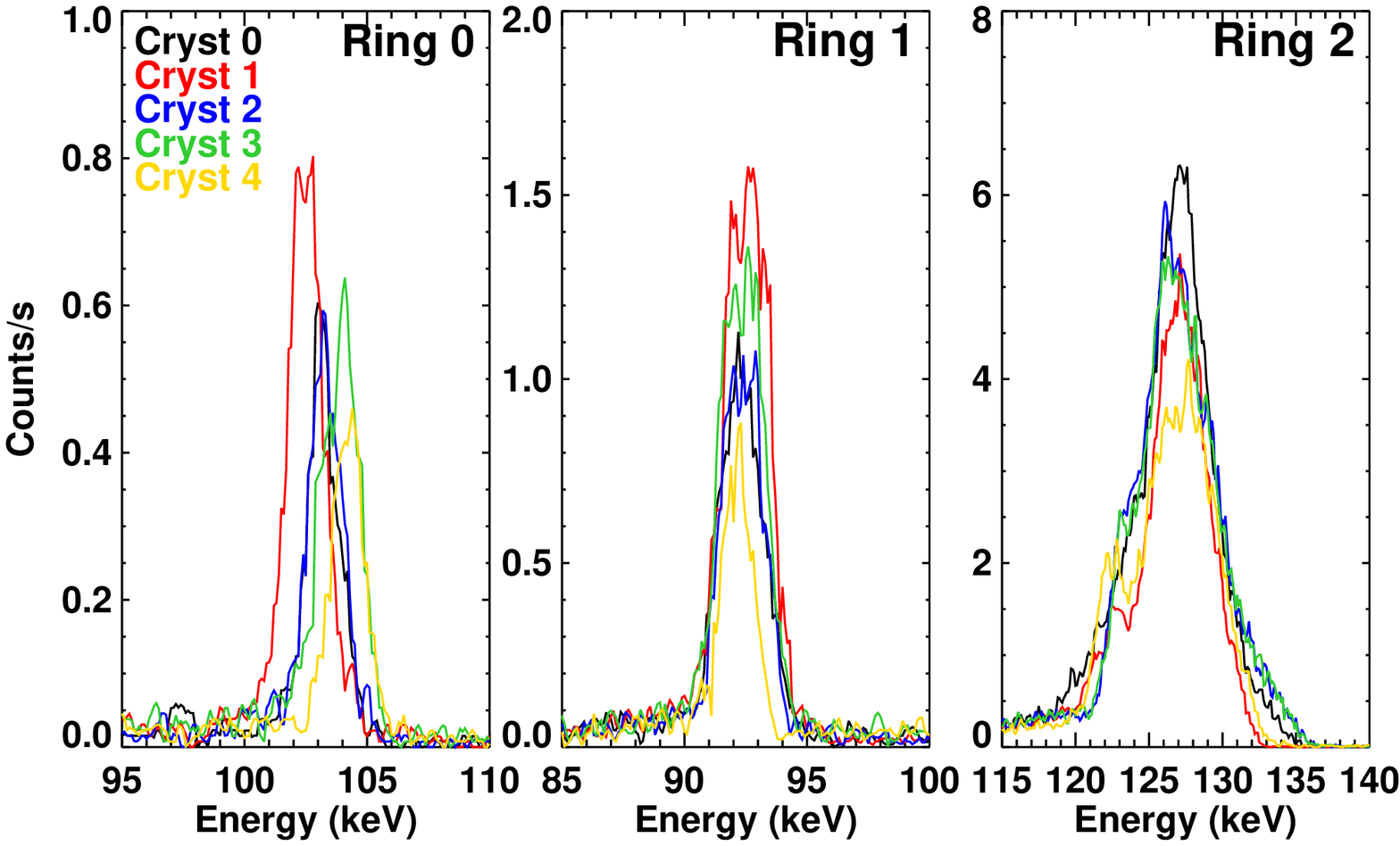}
\caption{Diffracted spectrum extracted for each crystal from the out-of-focus image (Figure~\ref{fig:PSFoutoffocus4m}). No background subtraction was performed, and no smoothing or binning was applied.}
\vspace{-1cm}
\label{fig:ringplots}
\end{center}
\end{figure}

The spectra diffracted by each individual crystal are shown in Figure~\ref{fig:ringplots}. These peaks are fit with a Gaussian function and the peak energies are converted to angular misalignment using the following formula:
\[
\left.
\begin{aligned}
{\Delta E \over E} = {\Delta \theta \over \theta} \; \\
2d_{hkl} \sin \theta = {hc \over E} \;
\end{aligned}
\right\}  \; \Delta \theta = {\Delta E \over E_{\rm goal}} \arcsin \left( {hc \over 2d_{hkl} E_{\rm goal}} \right) \\
\]
where the lower equation on the left hand side is the Bragg relation involving the d-spacing $d_{hkl}$ of the crystalline planes (defined by the Miller indices $h$, $k$ and $l$), $\theta$ is the incidence angle of the rays onto the planes, $h$ is Planck's constant and $c$ the speed of light. $E_{\rm goal}$ is the goal energy for a given ring (see Table \ref{tab:lensdetails}). The resulting angular misalignments are reported in Table \ref{tab:error_thT}. 

Our main goal was to be within the requirements for the Bragg angle standard deviation, $\sigma_{\theta T}$, which is by far the most constraining. While ring 0 is far from this goal, rings 1 and 2 are well within it. We showed in Ref. \cite{barriere.2011fk} that a standard deviation lower than $6''$ is possible with the glue we are using, and we confirm it again with this breadboard lens. The bad figure of ring 0 is likely due to an insufficient curing time. We left the crystal holder in position for only 4.5~h for the first ring, as opposed to 5.0~h to 5.25~h for the two next rings. Another explanation is the high packing factor, combined with the fact that we did not use a glue dispenser. The amount of glue injected was controlled by eye, and most of the time was overflowing on the next crystal's site. We learned as we progressed and improved the procedure for rings 1 and 2.

Although the low dispersion in orientation is an excellent result, we see that the offset of each ring exceeds by far the requirement. Since we glued crystals on small portion of rings, we can not distinguish between an angular offset of the crystals and a misorientation of the substrate. We nonetheless attribute these errors to the substrate orientation; while doing the lens alignment for the calibration measurements, we realized that a problem occurred with the rotating crystal method that led to significant misalignment of the substrate relative to the beam. 

 This problem is described in detail in section \ref{sec:pointingerror}. As a consequence, the three rings' optical axes are not well co-aligned. 

\begin{table}[htdp]
\begin{small}
\begin{center}
\begin{tabular}{|l|c|c|c|c|c||c|c|}
\hline
Ring & Cryst. 0 & Cryst. 1 & Cryst. 2 & Cryst. 3 & Cryst. 4 & $\Delta_{\theta T}$ & $\sigma_{\theta T}$ \\
\hline
\hline
0 & 52.2 $\pm$   0.5 & 24.6 $\pm$   0.4 & 55.0 $\pm$   0.5 & 77.0 $\pm$   0.5 & 91.6 $\pm$   0.5 & 60.1 & 25.7 \\
\hline
1 & -13.7 $\pm$   0.5 & -6.4 $\pm$   0.4 & -12.6 $\pm$   0.4 & -13.4 $\pm$   0.4 & -22.4 $\pm$   0.4 & -13.7 & 5.7 \\
\hline
2 & -19.2 $\pm$   0.3 & -17.8 $\pm$   0.3 & -27.8 $\pm$   0.5 & -24.1 $\pm$   0.5 & -16.2 $\pm$   0.4 & -21.0 & 4.8 \\
\hline
\end{tabular}
\end{center}
\end{small}
\caption{Angular misalignment about $\vec{\theta_T}$ in arc-seconds, and associated mean ($\Delta_{\theta T}$) and standard deviation ($\sigma_{\theta T}$) of the distribution.}
\label{tab:error_thT}
\end{table}%

\subsection{Errors on the two other angles}

After having entered the $\theta_T$ misalignments in the lens model, we use the position of the crystal footprints in the out-of-focus image to determine the $\theta_Z$ misalignments for each crystal, as reported in Table  \ref{tab:error_thZ}. Our ray trace model indicates that changing $\theta_Z$ by $10'$ leads to a vertical displacement (Y axis in Figure~\ref{fig:PSFoutoffocus4m}) of $\sim$0.5 mm, with no measurable energy change ($5 \times 10^{-5}$ keV). 
Our camera has a spatial resolution of 2~mm, which allows us to determine a crystal footprint with a precision of $\sim$0.5 mm (interpolating the intensity in each voxel), corresponding to a $\theta_Z$ misalignment of $\sim$$10'$. However, the quantum efficiency cross calibration between strips is estimated to be of the order of 20\%. So we did not go through a thorough determination of $\theta_Z$, and simply adjusted it by hand to have the contours of the simulated crystal footprints overlay the measured ones (green rectangles shown in Figure {\ref{fig:PSFoutoffocus4m}).

On the other hand, our ray trace model shows that the $\theta_R$ misalignment has almost no effect in this configuration; a crystal footprint moves of 70 $\mu$m/degree and the diffracted energy of 0.03 keV/degree. So we can not measure the $\theta_R$ misalignment here. Given that the crystals are glued with a bond line of $\sim$80 $\mu$m, we estimate that the error on $\theta_R$ can not exceed $27'$ (40$\mu m$ over 5 mm), and is most likely much smaller than this value. 

Most crystals have a $\theta_Z$ misalignment lower than $10'$, although the middle crystal of ring 1 is very poorly oriented, with an misalignment of about $-4500''$ (1.25$^{\circ}$). $\theta_Z$ was constrained by the two ledges at the tip of the crystal holder, relying on the external faces of the crystals (Figure~\ref{fig:photos}d). A careful mounting of the crystal in the holder seems sufficient to insure a standard deviation of the distribution lower than $10'$.

More than a quantitative result, we demonstrate here that it is possible to measure the angular misalignment of crystals using out-of-focus measurement with an imaging camera. An image intensifier with pixels of $\sim$0.5 mm, placed 6~m behind the lens, would yield arc-minute resolution.

\begin{table}[htdp]
\begin{small}
\begin{center}
\begin{tabular}{|l|c|c|c|c|c|}
\hline
Ring & Cryst. 0 & Cryst. 1 & Cryst. 2 & Cryst. 3 & Cryst. 4  \\
\hline
\hline
0 	& -8.3 & 0 & -8.3 & -8.3 & 8.3  \\
\hline
1 	& 0       & 0 & -75 & -25 & 0 \\
\hline
2	& 16.7 & -25 & 8.3 & -8.3 & 33.3 \\
\hline
\end{tabular}
\end{center}
\end{small}
\caption{Crystal misalignment about $\theta_Z$, in arc-minutes. The overlay of the simulated footprints with the measured ones was done by manually adjusting $\theta_Z$ for each crystal, by increment of $500''$ ($8.3'$). }
\label{tab:error_thZ}
\end{table}%

\subsection{In-focus measurements}
For the in-focus measurements, we placed the detector 1.5~m behind the lens. We acquired data with the source close to on-axis ($\theta_S=12''$, $\phi_S=90^{\circ}$) and $20'$ off-axis, and compared these data to simulations obtained with the lens model defined earlier.

Figure~\ref{fig:Diffrspc} shows the measured spectra and the simulated contribution of each ring for both source configurations. No background subtraction was performed, and no binning or smoothing was applied. 
There are two free parameters when fitting the spectrum diffracted by mosaic crystals: the mosaicity and the mean crystallite size (see e.g. \cite{halloin.2005gd}). We obtain the best fit with a crystal mosaicity of $8''$ and $120''$, and a crystallite size of 1 $\mu$m and 95 $\mu$m for the Si and Cu crystals, respectively. We note that our model (Darwin's model of mosaic crystals \cite{darwin.1922zg}) does not reproduce the large wings exhibited by the Cu crystals. This problem is well known and is currently being addressed as part of our study of crystals for Laue lenses \cite{Barriere:he5432}. Disregarding this point, our ray-trace code and the angular misalignments determined in the previous sections seem to provide a good modeling of the lens spectrum, even for an off-axis angle as large as $20'$.

Using the near on-axis run, we measure again the peak energy of each ring (Table \ref{tab:axisoffset}). Accounting for the fact that the source was $12''$ off axis, we find angular offsets in agreement with those determined using the out-of-focus acquisition.

Figure~\ref{fig:PSFatfocus} shows the images recorded by the detector with the source nearly on-axis and $20'$ off-axis. The contours of the simulated footprint of the crystals are overlaid in the off-axis case, showing, once again, that we have good agreement between simulation and data. One difference comes from the horizontal streaks produced by the wings of the Cu crystals. In the right panel, one can see the size of the focal spot produced by the 15-crystal prototype lens. 

\begin{figure}[p]
\begin{center}
\includegraphics[width=0.47\textwidth]{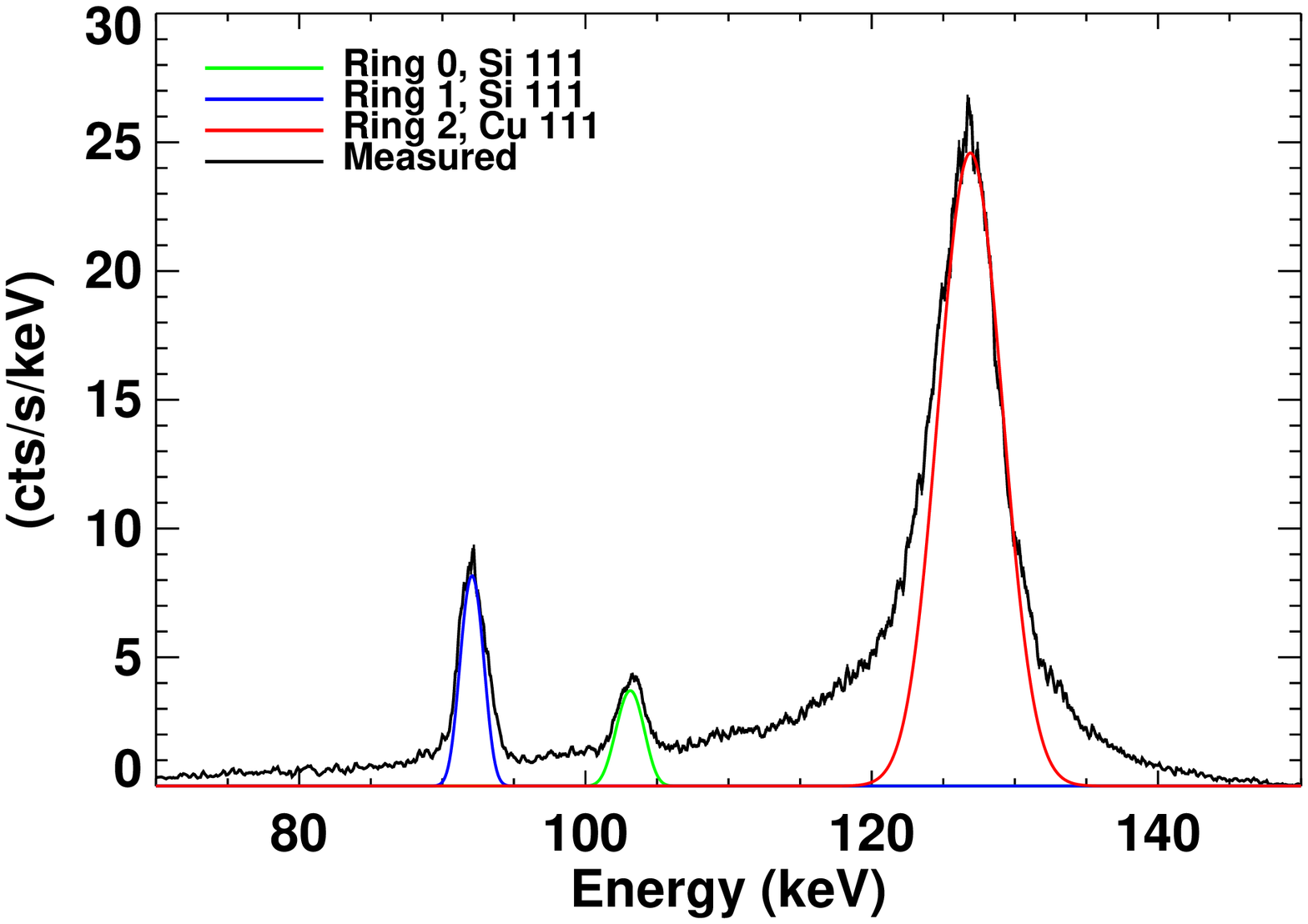}
\includegraphics[width=0.47\textwidth]{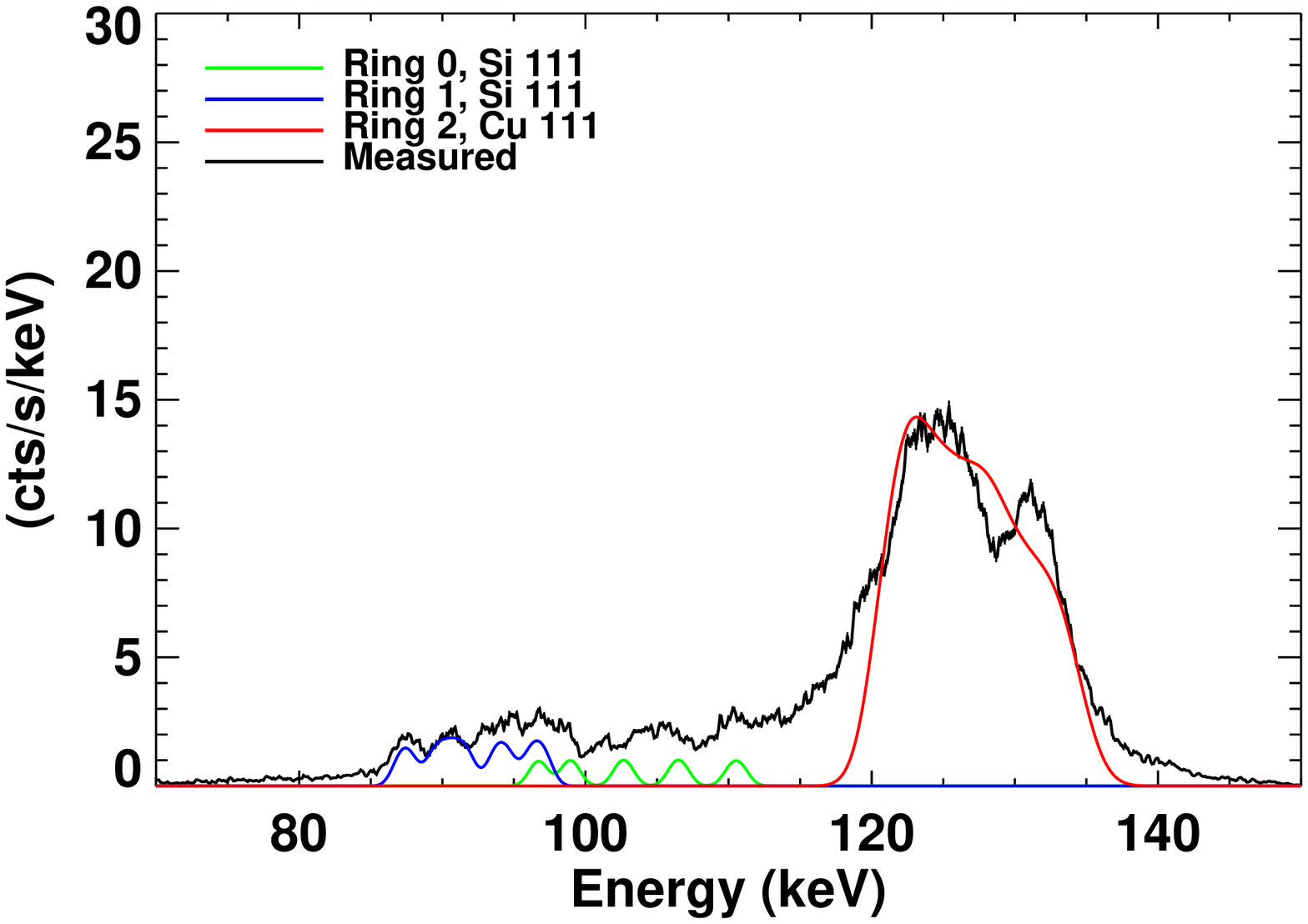}
\vspace{0.5cm}
\caption{Diffracted spectrum measured at the focus with the source nearly on-axis (left), and $20'$ off-axis (right). The simulated contribution of rings 0, 1 and 2 (green, blue and red lines, respectively) are shown as well.}
\vspace{-1cm}
\label{fig:Diffrspc}
\end{center}
\end{figure}

\begin{figure}[p]
\begin{center}
\includegraphics[width=0.47\textwidth]{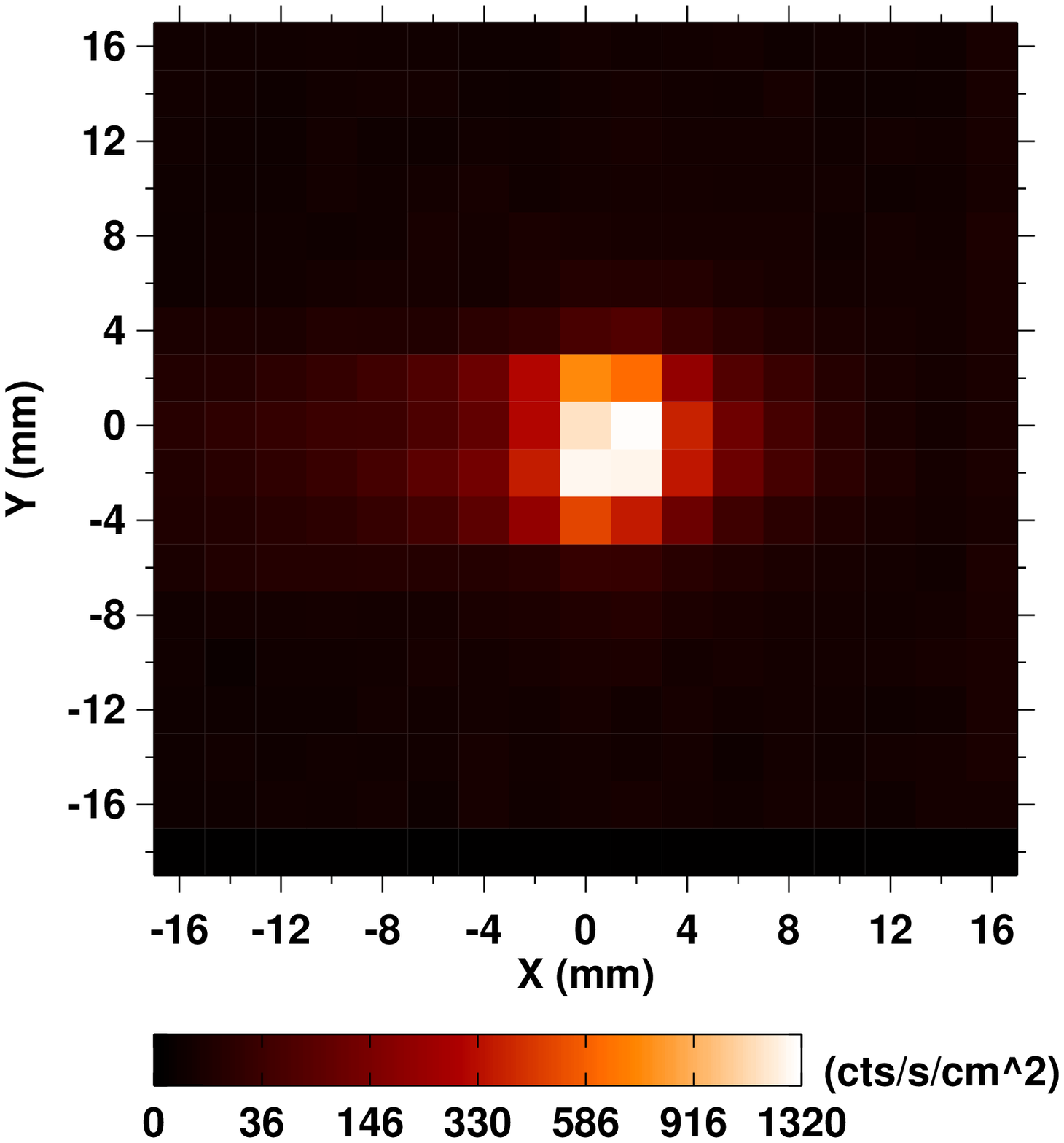}
\hspace{0.5 cm}
\includegraphics[width=0.47\textwidth]{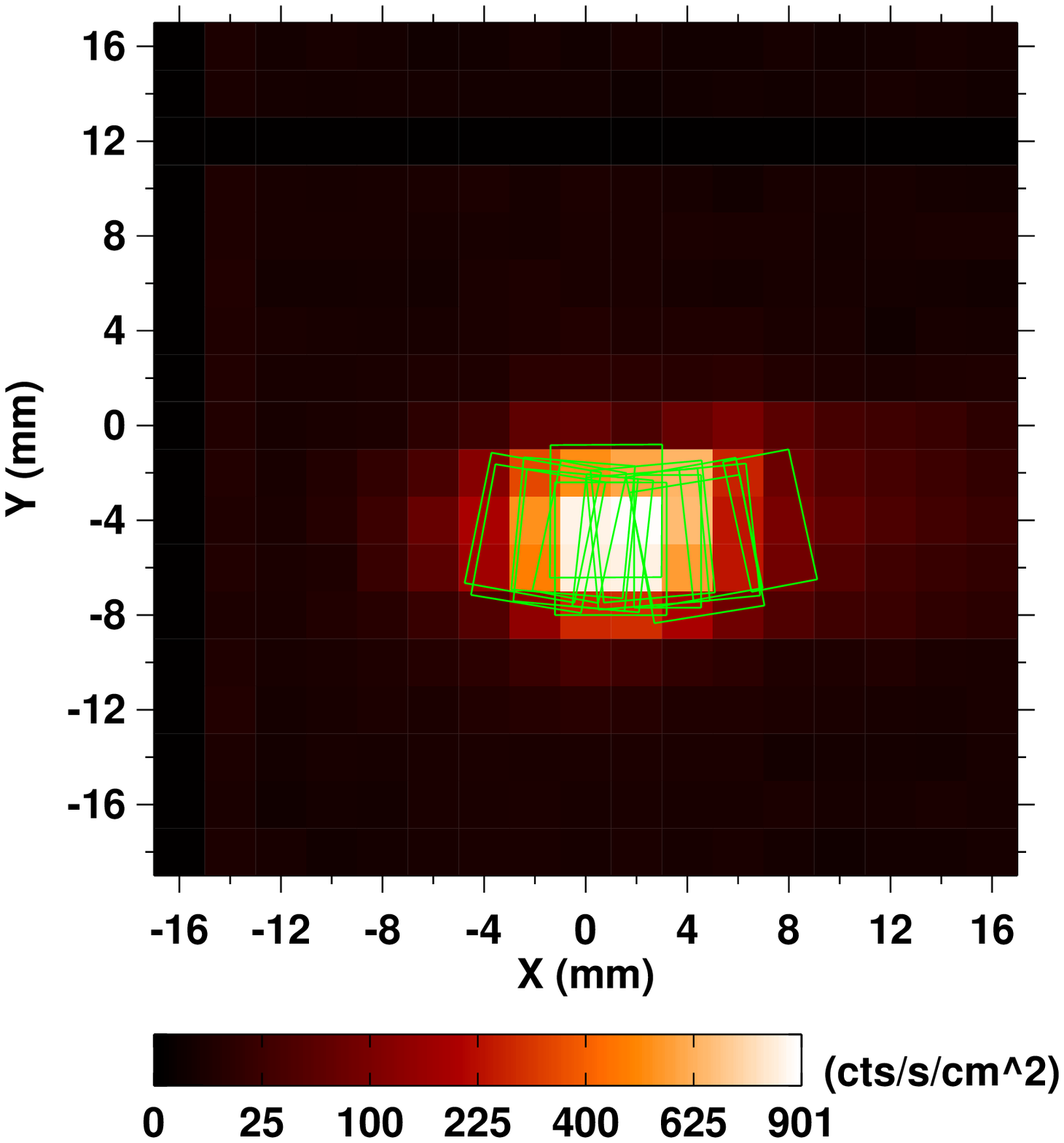}
\caption{Images acquired with the camera placed at the focus of the lens with the source nearly on-axis (left), and $20'$ off-axis (right). The simulated footprints of the 15 crystals comprising the lens are shown in the right panel. The horizontal streaks are due to the wings of the Cu crystals.}
\label{fig:PSFatfocus}
\end{center}
\end{figure}

\begin{table}[htdp]
\begin{small}
\begin{center}
\begin{tabular}{|c|c|c|c|}
\hline
Ring 	& Goal energy	& Measured energy 	& $\Delta_{\theta T}$ 	\\
\#		&  (keV)		& (keV) 			& ($''$)		\\
\hline
\hline
0 		&  101.878	& 103.177	$\pm$ 0.012	&  63.0 $\pm$ 2.5 \\
\hline
1		&  92.625		&   92.145	$\pm$ 0.005	&  -10.8 $\pm$ 2.3 \\
\hline
2		& 127.569 	&  126.776 $\pm$ 0.009	&  -17.8 $\pm$ 2.3 \\
\hline
\end{tabular}
\end{center}
\end{small}
\caption{Goal and measured peak energy for each ring, and derived angular offset ($\Delta_{theta T}$). \label{tab:axisoffset}}
\end{table}%

\subsection{Analysis of the pointing error}
\label{sec:pointingerror}

\begin{figure}[t]
\begin{center}
\includegraphics[width=0.6\textwidth]{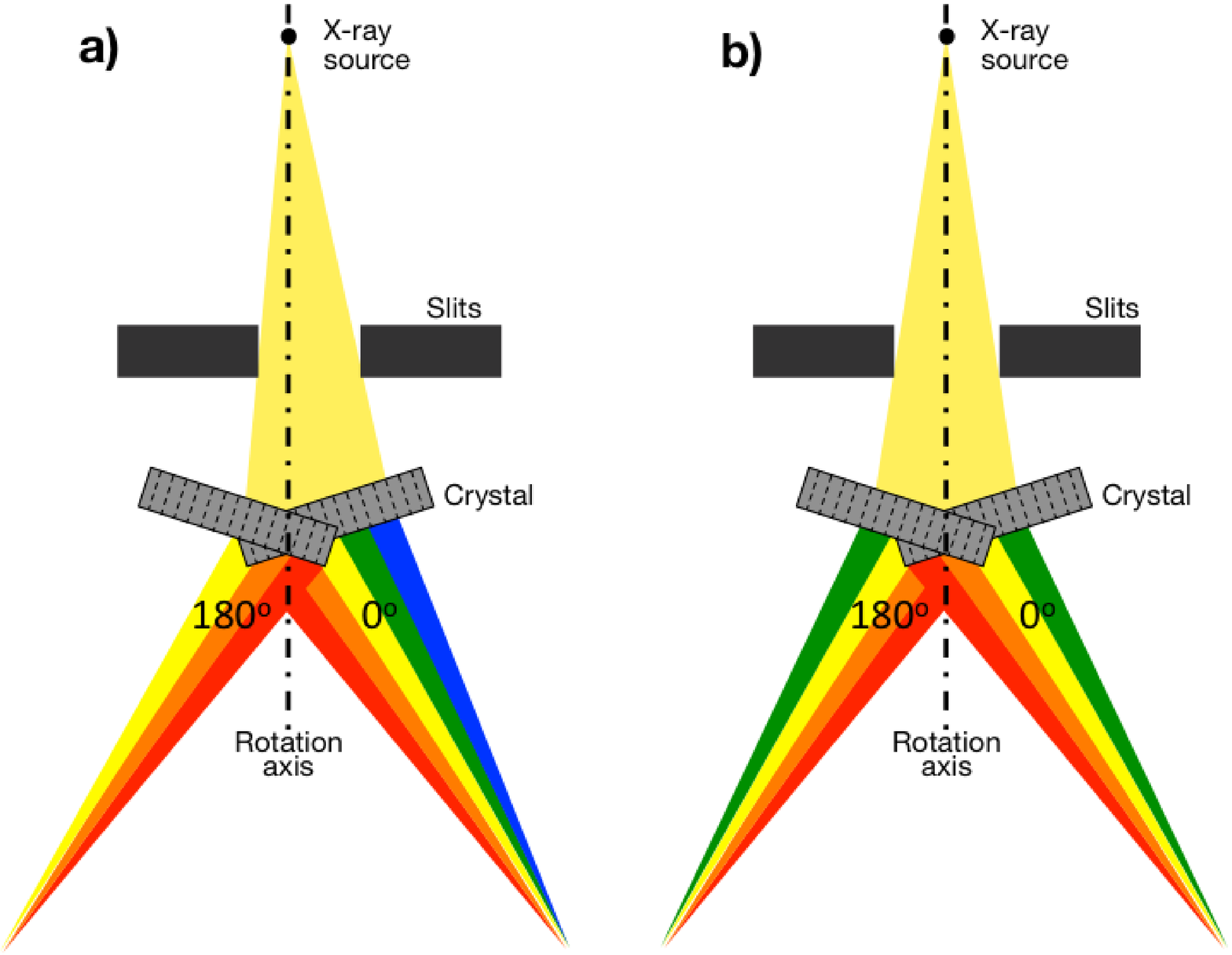}
\caption{Illustration of the artifact that occurred with the rotating crystal method. The crystal is represented by the grey rectangle, the dashed lines symbolizing the diffracting planes. In both a) and b) cases, the crystal is shown at two azimuthal angles of the lens, 180$^{\circ}$ apart, the dash-dot line representing the rotation axis. The colors used in the diffracted beam represent the energy, from low (red) to high (blue) energy.}
\label{fig:rotatingCryst}
\end{center}
\end{figure}

In this section, we analyze the cause of the angular offset between the crystal rings. The substrate was reoriented to point at the source before each ring was populated using the rotating crystal method (section \ref{sec:OrientSub}). While the method is potentially very accurate, our experimental setup introduced a bias. The problem was due to a misalignment between the slits defining the beam and the center of the lens, as illustrated in Figure~\ref{fig:rotatingCryst}.
With a diverging beam featuring a continuum spectrum, the angle of incidence (and therefore the energy diffracted) is position dependent. In Figure~\ref{fig:rotatingCryst} (panel a), although the rotation axis points at the source, the peak energy diffracted for the lens azimuthal angle of 180$^{\circ}$ is shifted towards low energy with respect to the peak energy diffracted at 0$^{\circ}$. This would appear as if the rotation axis was not pointing at the source. One can see in Figure~\ref{fig:rotatingCryst} (panel b) that the key point is to have the lens' center aligned with the center of the slits.

In the present case, the substrate positioning was done by eye (thanks to a laser that goes through the beamline and shows the beam) with an estimated precision of $\sim$1~mm, which results in an offset of the rotation axis of order of $16.5''$. This explains the offset observed for rings 1 and 2. For ring 0, our conclusion is that a mistake was made when the substrate was shifted from the rotating crystal position (center of the lens in the beam) to the gluing position (ring 0 in the beam) and then rotated to face the beam, which was done with a manual rotation-and-tilt stage.

It is clear from that experience that the orientation of the lens substrate is as sensitive as the orientation of the crystals themselves. The lens substrate should be mounted on a three-axis motorized stack of stages with a repeatability of the order of $1''$, and a better system should be used for monitoring the orientation of the substrate.

\section{Conclusions}
\label{sec:conclusions}
For more than two years a Laue lens assembly process has been under development at the Space Sciences Laboratory. It required the construction of an X-ray beamline and a dedicated end station, the so-called Laue lens assembly station. These tools and methods were tested with the realization of a breadboard Laue lens made of 10 silicon crystals and 5 copper crystals glued onto an aluminum substrate. Our goal is to meet the stringent requirements on crystal orientation accuracy imposed by the long focal length (30 m) necessary to build a LLT dedicated to the study of Type Ia supernovae.

The results of this first trial are very encouraging: considering ring 0 a trial run, we were able to quickly refine the assembly process and meet the requirement of $10''$ standard deviation on the Bragg angle misalignment for ring 1 and 2. The packing factor is as high as possible, with a nominal interspacing of 0.2 mm. However, the lens reference system showed its limits resulting in a poor co-alignment of each ring's optical axis. Although there are improvements to be made, this prototype shows that the criteria for building an efficient LLT for Type Ia SNe are within our reach.
 
The realization of a prototype also demonstrated our ability to simulate a Laue lens for the case of a source at finite distance, which is key to the assembly and calibration of any lens. Indeed, even a lens designed for sources at infinity would be assembled and calibrated in a diverging beam. A calibration procedure was developed and successfully applied to the characterization of the lens. 

The realization of this prototype served its purpose: test the tools and methods, and identify the critical points for further refinement. The LLAS is currently being upgraded in preparation for a second prototype assembly. The emphasis is on improving the thermal stability of the enclosure and the control of the substrate orientation with respect to the X-ray beam. We are also investigating alternative glues allowing the curing time to be significantly reduced. The next prototype will be composed of crystals optimized for diffraction at 120 keV, and the calibration procedure will have to determine the reflectivity in addition to the angular misalignments. We intend to put this next prototype through thermal-vacuum cycles and vibration tests, in order to move the Laue lens technology closer to technology readiness level 6, which would then allow it to be proposed for balloon-borne or satellite-borne missions.

\section*{Acknowledgments}       
Support for this work was provided by the Space Sciences Laboratory's Townes Fellowship in Experimental Astrophysics and Space Physics, and by NASA through the APRA grant NNG05WC28G. CW and LH acknowledges support from Science Foundation Ireland under research grant number 11-RFP-AST-3188. The authors are grateful to Nicolai F. Brejnholt for a thorough review of the manuscript, and to the anonymous referee for useful suggestions.


\bibliographystyle{elsarticle-num}
\bibliography{Lauelensproto.bib}







\end{document}